\documentclass[fleqn]{article}
\usepackage[LGR,T1]{fontenc}
\usepackage[margin=1.2in]{geometry} %
\usepackage{mathtools}
\usepackage{amssymb}
\usepackage{newtxtext}
\usepackage{newtxmath}
\usepackage{array}
\usepackage{booktabs}
\usepackage{mathrsfs}
\usepackage{dsfont}

\usepackage{cite}%
\usepackage{xcolor}
\usepackage{nicefrac}
\usepackage{wrapfig}
\usepackage{xspace}
\usepackage[small]{caption}
\usepackage{subcaption}  
\usepackage{braket}
\usepackage{mleftright}
\mleftright
\usepackage[symbol]{footmisc}
\usepackage{soul}
\usepackage[pdftex,hidelinks]{hyperref}
\usepackage[noabbrev]{cleveref}
\usepackage{orcidlink}
\usepackage[textsize=tiny,backgroundcolor=white]{todonotes}
\usepackage[normalem]{ulem}

\makeatletter
\g@addto@macro\bfseries{\boldmath}
\makeatother
\DeclareSymbolFont{upgreek}{LGR}{cmr}{m}{n}
\DeclareMathSymbol{\deltaup}{\mathord}{upgreek}{`d}
\DeclareMathSymbol{\piup}{\mathord}{upgreek}{`p}
\DeclareMathSymbol{\epsilonup}{\mathord}{upgreek}{`e}
\definecolor{c1}{RGB}{0,119,187}
\definecolor{c2}{RGB}{51,187,238}
\definecolor{c3}{RGB}{0,153,136}
\definecolor{c4}{RGB}{238,119,51}
\definecolor{c5}{RGB}{204,51,17}
\definecolor{c6}{RGB}{238,51,119}
\definecolor{c0}{RGB}{187,187,187}

\newcommand{\Group}[2]{\ensuremath{\text{#1}(#2)}}

\newcommand{\deltaCP}{\delta}
\DeclareMathOperator{\re}{Re}
\DeclareMathOperator{\im}{Im}

\DeclareMathOperator{\Tr}{Tr}
\DeclareMathOperator{\diag}{diag}

\newcommand{\dd}{\mathop{}\!\mathrm{d}}
\newcommand{\ii}{\hskip0.1ex\mathrm{i}\hskip0.1ex}

\newcommand{\CP}{\ensuremath{\mathcal{CP}}\xspace}
\newcommand{\ee}{\mathrm{e}}

\makeatletter
\newcommand*{\defeq}{\mathchoice{\mathrel{\rlap{%
\raisebox{0.24ex}{$\m@th\cdot$}}%
\raisebox{-0.24ex}{$\m@th\cdot$}}%
=}{\mathrel{\rlap{%
\raisebox{0.24ex}{$\m@th\cdot$}}%
\raisebox{-0.24ex}{$\m@th\cdot$}}%
=}{\mathrel{\rlap{%
\raisebox{0.08ex}{\small$\m@th\cdot$}}%
\raisebox{-0.28ex}{\small$\m@th\cdot$}}%
=}{\mathrel{\rlap{%
\raisebox{0.08ex}{\tiny$\m@th\cdot$}}%
\raisebox{-0.28ex}{\tiny$\m@th\cdot$}}%
=}}
\newcommand*{\eqdef}{\mathchoice{=\mathrel{\rlap{%
\raisebox{0.24ex}{$\m@th\cdot$}}%
\raisebox{-0.24ex}{$\m@th\cdot$}}}{%
=\mathrel{\rlap{%
\raisebox{0.24ex}{$\m@th\cdot$}}%
\raisebox{-0.24ex}{$\m@th\cdot$}}}{%
=\mathrel{\rlap{%
\raisebox{0.08ex}{\small$\m@th\cdot$}}%
\raisebox{-0.28ex}{\small$\m@th\cdot$}}}{%
=\mathrel{\rlap{%
\raisebox{0.08ex}{\tiny$\m@th\cdot$}}%
\raisebox{-0.28ex}{\tiny$\m@th\cdot$}}}%
}
\newcommand*{\transpose}{%
{\mathpalette\@transpose{}}%
}
\newcommand*{\@transpose}[2]{%
\raisebox{\depth}{$\m@th#1\intercal$}%
}
\usepackage{acro}

\DeclareAcronym{2HDM}{
  short = 2HDM,
  long = two-Higgs doublet model,
  pdfcomment = two-Higgs doublet model
}

\DeclareAcronym{BSM}{
  short = BSM,
  long = beyond the standard model,
  pdfcomment = beyond the standard model
}

\DeclareAcronym{BU}{
  short = BU,
  long = bottom-up,
  pdfcomment = bottom-up
}

\DeclareAcronym{CG}{
  short = CG,
  long = Clebsch--Gordan,
  pdfcomment = Clebsch--Gordan
}

\DeclareAcronym{EFT}{
  short = EFT,
  long = effective field theory,
  pdfcomment = effective field theory
}

\DeclareAcronym{FCNC}{
  short = FCNC,
  long = flavor changing neutral current,
  pdfcomment = flavor changing neutral current
}

\DeclareAcronym{FI}{
  short = FI,
  long = Fayet--Iliopoulos \cite{Fayet:1974jb},
  pdfcomment = Fayet--Iliopoulos \cite{Fayet:1974jb}
}

\DeclareAcronym{GS}{
  short = GS,
  long = Green--Schwarz \cite{Green:1984sg},
  pdfcomment = Green--Schwarz \cite{Green:1984sg}
}

\DeclareAcronym{GUT}{
  short = GUT,
  long = Grand Unified Theory,
  pdfcomment = Grand Unified Theory
}

\DeclareAcronym{IO}{
  short = IO,
  long = inverted ordering,
  pdfcomment = inverted ordering
}

\DeclareAcronym{IR}{
  short = IR,
  long = infrared,
  pdfcomment = infrared
}

\DeclareAcronym{LEET}{
  short = LEET,
  long = low-energy effective theory,
  pdfcomment = low-energy effective theory
}

\DeclareAcronym{LHC}{
  short = LHC,
  long = Large Hadron Collider,
  pdfcomment = Large Hadron Collider
}

\DeclareAcronym{MIHO}{
  short = MIHO,
  long = modular invariant holomorphic observables,
  pdfcomment = modular invariant holomorphic observables
}

\DeclareAcronym{MSSM}{
  short = MSSM,
  long = minimal supersymmetric standard model,
  pdfcomment = minimal supersymmetric standard model
}

\DeclareAcronym{NO}{
  short = NO,
  long = normal ordering,
  pdfcomment = normal ordering
}

\DeclareAcronym{OPE}{
  short = OPE,
  long = operator product expansion,
  pdfcomment = operator product expansion
}

\DeclareAcronym{QFT}{
  short = QFT,
  long = quantum field theory,
  pdfcomment = quantum field theory
}

\DeclareAcronym{RG}{
  short = RG,
  long = renormalization group,
  pdfcomment = renormalization group
}

\DeclareAcronym{RGE}{
  short = RGE,
  long = renormalization group equation,
  pdfcomment = renormalization group equation
}

\DeclareAcronym{SM}{
  short = SM,
  long = standard model,
  pdfcomment = standard model
}

\DeclareAcronym{SUSY}{
  short = SUSY,
  long = supersymmetry,
  pdfcomment = supersymmetry
}

\DeclareAcronym{TD}{
  short = TD,
  long = top-down,
  pdfcomment = top-down
}

\DeclareAcronym{UV}{
  short = UV,
  long = ultraviolet,
  pdfcomment = ultraviolet
}

\DeclareAcronym{VEV}{
  short = VEV,
  long = vacuum expectation value,
  pdfcomment = vacuum expectation value
}

\DeclareAcronym{VVMF}{
  short = VVMF,
  long = vector-valued modular form,
  pdfcomment = vector-valued modular form
}
\newcommand*{\mytitle}{Scale-independent relations between neutrino mass parameters}
\title{\mytitle}

\hypersetup{
    pdftitle = {\mytitle},
    pdfauthor = {Mu-Chun Chen, Shaheed Perez, Michael Ratz}
}

\newcommand{\orcidauthorA}{0000-0002-5749-2566} 
\newcommand{\orcidauthorB}{0009-0001-7955-3681} 
\newcommand{\orcidauthorC}{0000-0003-2512-7422}

\begin{document}
\begin{center}
{\Large\sffamily\bfseries\mytitle}

\vspace{1cm}

\renewcommand*{\thefootnote}{\fnsymbol{footnote}}

\textbf{%
Mu-Chun Chen\rlap{,}\footnote{muchunc@uci.edu}{}\textsuperscript{,\orcidlink{\orcidauthorA}} 
Shaheed Perez\rlap{,}\footnote{shaheedp@uci.edu}{}\textsuperscript{,\orcidlink{\orcidauthorB}} and 
Michael Ratz\footnote{mratz@uci.edu}\textsuperscript{,\orcidlink{\orcidauthorC}}}
\\[8mm]
\textit{\small~Department of Physics and Astronomy, University of California, Irvine, CA 92697-4575 USA}
\end{center}
\begin{abstract}
  Theories of flavor operate at various scales. 
  Recently it has been pointed out that in the context of modular flavor symmetries certain combinations of observables are highly constrained, or even uniquely fixed, by modular invariance and holomorphicity. 
  We find that even in the absence of supersymmetry these combinations are surprisingly immune against quantum corrections.
  This applies, in particular, to the \ac{SM} and certain \acp{2HDM}.
\end{abstract}

\acresetall
\section{Introduction}
\label{sec:Introduction}

Theories of flavor accommodate, or even predict, fermion masses, mixing angles and $\CP$ phases, which constitute a significant fraction of the \ac{SM} parameters. 
The scale of new physics underlying the corresponding models, which we will denote by $\Lambda_\mathrm{flavor}$, generally are different from scales at which experimental measurements are made. 
This means that quantum corrections to the model predictions have to be taken into account. 
This raises the question of whether there are predictions that do not depend on the scale $\Lambda_\mathrm{flavor}$ at which the model is defined.  

In the context of modular flavor symmetries \cite{Feruglio:2017spp} (for reviews see e.g.~\cite{Feruglio:2019ybq,Almumin:2022rml,Kobayashi:2023zzc,Ding:2023htn,Ding:2024ozt,Feruglio:2025ztj}) it has recently been pointed out that there are certain combinations of entries of the Weinberg operator are independent of the modulus $\tau$~\cite{Chen:2024otk}. 
In addition, these combinations are known to be \ac{RG} invariant at 1-loop~\cite{Chang:2002yr}. 
This latter statement holds both in \ac{SM} and \ac{MSSM}. 

The purpose of this analysis is to discuss the impact of quantum corrections on the above invariance in the absence of \ac{SUSY}. 
This is motivated also by the recent proposal of non-holomorphic modular flavor symmetries \cite{Qu:2024rns,Ding:2024inn,Qu:2025ddz}, in non-supersymmetric setups.

\section{Neutrino masses described by the Weinberg operator}

We consider scenarios in which neutrino masses are described by the Weinberg operator. 
In the supersymmetric context, the superpotential of the lepton sector is then given by
\begin{equation}\label{eq:W_lepton_mass}
  \mathscr{W}_\mathrm{lepton~mass}
  =
  Y_e^{gf}\,L_g\cdot H_d\,E_f
  +
  \frac{1}{2}\kappa_{gf}\,L_g\cdot H_u\,L_f\cdot H_u\;.
\end{equation}
Here, the superfields $L^f$ and $E^f$ denote the three generations of the $\Group{SU}{2}_\mathrm{L}$ charged lepton doublets and singlets, respectively. 
The flavor indices $f$ and $g$. 
The superfields $H_{u/d}$ stand for the \ac{MSSM} Higgs doublets.
In \eqref{eq:W_lepton_mass}, ``$\cdot$'' indicate contractions with the Levi--Civita symbol.
$m_\nu=v_u^2\,\kappa$ is the neutrino mass matrix, with  $\kappa$ being the effective neutrino mass operator.
Finally, $Y_e$ denotes the charged lepton Yukawa couplings. 
In models based on modular flavor symmetries, $\kappa$ and $Y_{e}$ are given in terms of the modular forms.

In the \ac{SM} amended by the Weinberg operator, the lepton masses are described by 
\begin{equation}\label{eq:L_lepton_mass}
    \mathscr{L}_\mathrm{lepton~mass}=-Y_e^{gf}\,\overline{\ell_{\mathrm{L},g}}e_{\mathrm{R}\,g}\cdot\phi-\frac{1}{4}\kappa_{gf}\ell^g\cdot\phi\,\ell^f\cdot\phi
 +\text{h.c.}\;.
\end{equation}
Here, $\ell_{\mathrm{L},f}$ denote the lepton doublets, $e_{\mathrm{R}\,g}$ the right-handed charged leptons, and $\phi$ the \ac{SM} Higgs. 

Apart from the charged lepton masses, $m_f=y_f\,v_\mathrm{EW}$ with $v_\mathrm{EW}$ denoting the \ac{VEV} of the electroweak Higgs $\phi$, the lepton sector has 9 flavor parameters, 
\begin{equation}\label{eq:flavor_parameters}
    \{\xi_i\}=\{m_1,m_2,m_3,\theta_{12},\theta_{13},\theta_{23},\deltaCP,\varphi_1,\varphi_2\}\;.
\end{equation}
Out of these parameters, two mass squared differences, $\Delta m_{ij}^2\defeq m_i^2-m_j^2$, and the mixing angles $\theta_{ij}$ have been measured with relatively good precision, see e.g.~\cite{Esteban:2024eli}. 
The recent JUNO results \cite{JUNO:2025gmd} have significantly reduced the error bars of $\theta_{12}$ and $\Delta m_{12}^2\defeq m_2^2-m_1^2$, see e.g.\ \cite{Capozzi:2025ovi}. 
On the other hand, the absolute neutrino mass scale and the Dirac phase $\deltaCP$ are subject to constraints but not determined precisely. 
We currently do not know whether neutrinos are Majorana fermions, and thus have no knowledge of the values of the Majorana phases $\varphi_i$. 

\section{Lepton flavor parameters and quantum corrections}

\Cref{eq:W_lepton_mass,eq:L_lepton_mass} contain the Weinberg operator,
\begin{equation}
 \mathscr{L}_\kappa=-\frac{1}{4}\kappa_{gf}\ell^g\cdot\phi\,\ell^f\cdot\phi
 +\text{h.c.}\;.
\end{equation}
$\kappa$ is a symmetric matrix of mass dimension $-1$. 

Throughout this study, we will work in a basis in which the fields are canonically normalized and $Y_e$ is diagonal and positive, 
\begin{equation}\label{eq:Y_e_diag}
    Y_e=\diag(y_e,y_\mu,y_\tau)\quad\text{with }y_f>0\text{ for }f\in\{e,\mu,\tau\}\;.
\end{equation}
In this basis, all the renormalizable interactions in the lepton sector are diagonal in flavor space. 

\subsection{Invariants}

In the basis chosen as given in \eqref{eq:Y_e_diag}, we define the invariants
\begin{equation}\label{eq:definition_of_I_fg}
 I_{fg}\defeq \frac{(m_\nu)_{ff}\,(m_\nu)_{gg}}{\bigl((m_\nu)_{fg}\bigr)^2}  
 =
 \frac{\kappa_{ff} \,\kappa_{gg}}{\bigl(\kappa_{fg} \bigr)^2}  
 \;,
\end{equation}
where no summation over the flavor indices $f$ and $g$ is implied. We are interested in quantum corrections to these combinations. 
In order to obtain the second equality in  \eqref{eq:definition_of_I_fg}, we have to assume that the normalizations of the three lepton doublets coincide at a given scale.The focus of this study is on the \ac{RG} stability of the $I_{fg}$ \eqref{eq:definition_of_I_fg}. 

A key feature of these expressions is that they can be entirely expressed in terms of observable flavor parameters. 
Explicitly,
\begin{subequations}
 \begin{align}
    \label{eq:simpFormOfI12} I_{12} &= \frac{a_0\left[\widetilde{m}_1(c_{23}s_{12} + \ee^{-\ii\deltaCP}c_{12}s_{13}s_{23})^2 + \widetilde{m}_2(c_{12}c_{23} - \ee^{-\ii\deltaCP}s_{12}s_{13}s_{23})^2 + m_3c_{13}^2s_{23}^2 \right]}{\left[\widetilde{m}_1c_{12}(c_{23}s_{12} + \ee^{-\ii\deltaCP}c_{12}s_{13}s_{23}) - \widetilde{m}_2s_{12}(c_{12}c_{23} - \ee^{-\ii\deltaCP}s_{12}s_{13}s_{23}) - \ee^{\ii\deltaCP}m_3s_{13}s_{23} \right]^2} \\[1em]
    \label{eq:simpFormOfI13} I_{13} &= \frac{a_0\left[\widetilde{m}_1(\ee^{-\ii\deltaCP}c_{12}c_{23}s_{13} - s_{12}s_{23})^2 + \widetilde{m}_2(\ee^{-\ii\deltaCP}c_{23}s_{12}s_{13} + c_{12}s_{23})^2 + m_3c_{13}^2c_{23}^2 \right]}{\left[\widetilde{m}_1c_{12}(\ee^{-\ii\deltaCP}c_{12}c_{23}s_{13} - s_{12}s_{23}) + \widetilde{m}_2s_{12}(\ee^{-\ii\deltaCP}c_{23}s_{12}s_{13} + c_{12}s_{23}) - \ee^{\ii\deltaCP}m_3c_{13}c_{23} \right]^2} \\[1em]
    \label{eq:simpFormOfI23} I_{23} &=  \left[m_3 c^2_{13} s^2_{23}+\widetilde{m}_1 \left(c _{23} s _{12}+ \ee^{-\ii \deltaCP } c _{12} s _{13} s _{23}\right)^2+\widetilde{m}_2 \left(c _{12} c _{23}- \ee^{-\ii \deltaCP } s _{12} s _{13} s _{23}\right)^2\right]
    \nonumber\\
    &\qquad  {}\times
    \frac{4 \left[ m_3 c^2_{13} c^2_{23}+\widetilde{m}_2 
    \left(\ee^{-\ii \deltaCP } c_{23} s_{12} s_{13}+ c_{12} s _{23}\right)^2+\widetilde{m}_1
    \left(\ee^{-\ii \deltaCP }c_{12} c_{23} s_{13}- s_{12} s _{23}\right)^2\right]}{%
    \left[
        (\widetilde{m}_1 a_1
        +\widetilde{m}_2 a_2)
        -m_3 \sin (2 \theta _{23}) c^2_{13}
        \right]^2}
 \;,
 \end{align}
\end{subequations}
$s_{ij}\defeq\sin\theta_{ij}$, $c_{ij}\defeq \cos\theta_{ij}$, $t_{ij}\defeq \tan\theta_{ij}$, and
\begin{subequations}
    \begin{align}
    \label{eq:newDefOfa0} a_0 &\defeq \widetilde{m}_1c_{12}^2 + \widetilde{m}_2s_{12}^2 + \ee^{2\ii\deltaCP}m_3t_{13}^2 \\[1em]
    \label{eq:newDefOfa1}a_1 &\defeq \left[\left(s^2_{12}-\ee^{-2\ii \deltaCP }c^2_{12} s^2_{13}\right) 
    \sin (2 \theta _{23})-\ee^{-\ii \deltaCP } \cos (2 \theta _{23}) \sin (2 \theta_{12})
    s_{13}\right]
    \;,\\[1em]
    \label{eq:newDefOfa2}a_2 &\defeq 
    \left[\ee^{-\ii \deltaCP} \cos(2 \theta_{23}) 
    \sin (2 \theta _{12}) s_{13}+\left( c ^2_{12}- \ee^{-2\ii \deltaCP }s^2_{12} 
    s^2_{13}\right) \sin (2 \theta _{23})\right]\;.
\end{align}
\end{subequations}
The invariants $I_{fg}$ depend on $m_1$, $m_2$, $\varphi_1$ and $\varphi_2$ only via the combinations $\widetilde{m}_1\defeq m_1\,\ee^{\ii\varphi_1}$ and $\widetilde{m}_2\defeq m_2\,\ee^{\ii\varphi_2}$.
As $I_{fg}$ are complex, each of them contains two real flavor parameters. 
This means that, unless there are degeneracies, 6 independent linear combinations out of the 9 flavor parameters $\xi_i$ in \eqref{eq:flavor_parameters} are described by the $I_{fg}$. 
As discussed in \cite{Chen:2024otk}, zeros and poles of the $I_{fg}$ correspond to texture zeros of the Weinberg operator. 

Why are we interested in these invariants, $I_{fg}$? There are two main reasons. 
First of all, they are \ac{RG} invariant at the 1-loop level \cite{Chang:2002yr}, as we shall discuss in more detail in \Cref{sec:RGEs}. 
Additionally, they turn out to have remarkable properties in the framework of modular flavor symmetries. 
For instance, in the Feruglio model \cite{Feruglio:2017spp}, $I_{12}=-2$ and $I_{13}\,I_{23}=-32$, independently of the value of the modulus \cite{Chen:2024otk}. 
That is, these invariants carry a large amount of the information on the modular symmetries. 
As we shall see next, they are not only independent of the modulus but also, for all practical purposes, insensitive to the definition flavor scale $\Lambda_\mathrm{flavor}$ of the model.

\subsection{Renormalization group equations}
\label{sec:RGEs}

In~\cite{Chang:2002yr} it has been found that $I_{fg}$ defined in \eqref{eq:definition_of_I_fg} are independent of the renormalization scale at one-loop. 
In the supersymmetric context, one may view this as being a simple consequence of the non-renormalization theorem. 
Only the wave-function renormalization constants depend on the scale, and the latter cancel in the $I_{fg}$ expressions~\cite{Haba:1999ca}. 
The \ac{RG}-invariance of $I_{fg}$ thus holds at all loop levels in supersymmetric models. 
So in the following, we will focus on non-supersymmetric case.

Since $\kappa$ is a symmetric matrix, its \ac{RGE} has the form 
\begin{equation}\label{eq:newPropFormOfRGE}
    \frac{\dd}{\dd t}\kappa =\sum_k \kappa^{(k)}\defeq \sum_k \left(\frac{1}{16\piup^2}\right)^k\, \left[\alpha^{(k)}\kappa + P^{(k)}\, \kappa + \kappa \bigl(P^{(k)}\bigr)^{\transpose} + Q^{(k)} \,\kappa\, \bigl(Q^{(k)}\bigr)^{\transpose}\right] \;.
\end{equation}
Here, the superscript ``$(k)$'' indicates the loop order, $\alpha^{(k)}$ denotes the scalar contribution, which includes gauge couplings and parameters of the Higgs potential, and $P^{(k)}$ and $Q^{(k)}$ are implicitly defined by \eqref{eq:newPropFormOfRGE}. 
We will provide explicit expressions below in \eqref{eq:PandQMatrixAt2Loop}. 
The $t$-derivative is the logarithmic derivative with respect to the renormalization scale $\mu$,
\begin{equation}\label{eq:defOfdt}
  \frac{\dd}{\dd t}\defeq \mu\frac{\dd}{\dd\mu} \;,
\end{equation}
i.e.\ $t=\ln(\mu/\mu_0)$ with some reference scale $\mu_0$. 
In \eqref{eq:newPropFormOfRGE}, $k$ indicates the loop level, and the $\alpha^{(k)}$ are flavor-independent coefficients. 
The matrices $P^{(k)}$, $Q^{(k)}$ are composed of the renormalizable couplings of the theory and diagonal,
\begin{subequations}\label{eq:P-and_Q-matrices_1}
\begin{align}
  P^{(k)}&=\diag\bigl(P^{(k)}_1,P^{(k)}_2,P^{(k)}_3\bigr)\;,\\
  Q^{(k)}&=\diag\bigl(Q^{(k)}_1,Q^{(k)}_2,Q^{(k)}_3\bigr)\;.  
\end{align}  
\end{subequations} 
At one-loop, $P^{(1)}=C_e\,Y_e Y_e^{\dagger}=C_e\,\diag(y_e^2,y_\mu^2,y_\tau^2)$ with $Y_e$ being the charged lepton Yukawa matrix \eqref{eq:Y_e_diag}.  
$C_e=-3/2$ in the \ac{SM}~\cite{Antusch:2001ck} and two-Higgs models~\cite{Antusch:2001vn}, and $C_e=1$ in the \ac{MSSM}~\cite{Chankowski:1993tx,Babu:1993qv} (see e.g.\ \cite{Xing:2007fb,Xing:2020ijf} for reviews). 
At the 1-loop level in \eqref{eq:newPropFormOfRGE}, there is only one matrix in flavor space, $P^{(1)}$, and we can choose $Q^{(1)}=\mathds{1}$. 

In order to have nonzero \ac{RG} effects, we consider two-loop \acp{RGE}~\cite{Machacek:1984zw,Machacek:1983fi,Luo:2002ti,Schienbein:2018fsw}. 
The two-loop contribution in \eqref{eq:newPropFormOfRGE} in the \ac{SM} is given by~\cite{Ibarra:2024tpt}
\begin{align}
    \kappa^{(2)}
    &= \left(\frac{1}{16\piup^2}\right)^2 \left[\alpha^{(2)}\,\kappa + P^{(2)}\, \kappa + \kappa\, \bigl(P^{(2)}\bigr)^{\transpose} + Q^{(2)}\, \kappa\, \bigl(Q^{(2)}\bigr)^{\transpose}\right] \; ,\label{eq:2ndLoopNewFormOfRGE} 
\end{align}
where now there is a nontrivial $Q$-matrix,
\begin{subequations}\label{eq:PandQMatrixAt2Loop}    
\begin{align}
     P^{(2)} &= \left(-\frac{57}{16}g_1^2 + \frac{33}{16}g_2^2 + \frac{5}{4}T\right)\,Y_eY_e^{\dagger} + \frac{19}{4}Y_eY_e^{\dagger}\,Y_eY_e^{\dagger} \; ,\label{eq:PMatrixAt2Loop} \\
     Q^{(2)} &= \sqrt{2}\,Y_eY_e^{\dagger} \; .\label{eq:QMatrixAt2Loop}
 \end{align}
\end{subequations}
Here, $g_1$ and $g_2$ are the running gauge coupling constants and $T \defeq \Tr[Y_eY_e^{\dagger} + 3Y_uY_u^{\dagger} + 3Y_dY_d^{\dagger}]$, with $Y_u$ and $Y_d$ being the Yukawa coupling matrices for the up-type quarks and the down-type quarks, respectively. 

Analogously to \eqref{eq:newPropFormOfRGE}, we can write the loop expansion of the $I_{fg}$ as 
\begin{align}
    \dot I_{fg} &\defeq 
     \frac{\dd}{\dd t}I_{fg} = \sum_k I_{fg}^{(k)} \;,\label{eq:I_fgLoopExpansion}
\end{align}
where
\begin{align}     
     I_{fg}^{(k)} &= \frac{\kappa_{ff}^{(k)}\,\kappa_{gg}}{\bigl(\kappa_{fg}\bigr)^2} + \frac{\kappa_{ff}\,\kappa_{gg}^{(k)}}{\bigl(\kappa_{fg}\bigr)^2} - 2\frac{\kappa_{ff}\,\kappa_{gg}}{\bigl(\kappa_{fg}\bigr)^3}\,\kappa_{fg}^{(k)}\;.\label{eq:I_fgAtk}
\end{align}
Truncating \eqref{eq:newPropFormOfRGE} at two-loop level, i.e.\ $k=2$, and inserting this truncation into \eqref{eq:I_fgLoopExpansion} we obtain 
\begin{align}
\frac{\dd}{\dd t}I_{fg}
&= \frac{(\kappa_{ff}^{(1)}+\kappa_{ff}^{(2)})\,\kappa_{gg}}{\kappa_{fg}^2} + \frac{\kappa_{ff}\,(\kappa_{gg}^{(1)}+\kappa_{gg}^{(2)})}{\kappa_{fg}^2} -2\frac{\kappa_{ff}\,\kappa_{gg}}{\kappa_{fg}^3}\,(\kappa_{fg}^{(1)}+\kappa_{fg}^{(2)}) \notag\\[1em]
    &= \left[\frac{\kappa_{ff}^{(1)}\,\kappa_{gg}}{\kappa_{fg}^2} + \frac{\kappa_{ff}\,\kappa_{gg}^{(1)}}{\kappa_{fg}^2} -2\frac{\kappa_{ff}\,\kappa_{gg}}{\kappa_{fg}^3}\,\kappa_{fg}^{(1)}\right] + \left[\frac{\kappa_{ff}^{(2)}\,\kappa_{gg}}{\kappa_{fg}^2} + \frac{\kappa_{ff}\,\kappa_{gg}^{(2)}}{\kappa_{fg}^2} -2\frac{\kappa_{ff}\,\kappa_{gg}}{\kappa_{fg}^3}\,\kappa_{fg}^{(2)}\right] \notag\\[1em]
    &\eqdef I_{fg}^{(1)} + I_{fg}^{(2)}\;.
\end{align}
Then, we can calculate $I_{fg}^{(k)}$ as
\begin{align}
     I_{fg}^{(k)} &=  \frac{\kappa_{ff}\,\kappa_{gg}}{(16\piup^2)^k\kappa_{fg}^2}\left[\alpha^{(k)} + 2P_{ff}^{(k)} + \bigl(Q_{ff}^{(k)}\bigr)^2 + \alpha^{(k)} + 2P_{gg}^{(k)} + \bigl(Q_{gg}^{(k)}\bigr)^2 
    \right. \nonumber\\
     &\hphantom{{}=\frac{\kappa_{ff}\,\kappa_{gg}}{(16\piup^2)^k\kappa_{fg}^2}}\quad{}\left.\vphantom{\bigl(Q_{gg}^{(k)}\bigr)^2} 
    -2\bigl(\alpha^{(k)} + P_{ff}^{(k)}  +  P_{gg}^{(k)} + Q_{ff}^{(k)}  Q_{gg}^{(k)}\bigr)\right] \nonumber\\
    &= \frac{\kappa_{ff}\,\kappa_{gg}}{(16\piup^2)^k\kappa_{fg}^2} \left(Q_{ff}^{(k)} -Q_{gg}^{(k)} \right)^2 \; .\label{eq:FinalResultForDerOfI_inNewNot}
\end{align}
This shows that the $\alpha$- and $P$-type terms in \eqref{eq:newPropFormOfRGE} do not affect the \acp{RGE} of the invariants. 
At 1-loop, $Q_{ff}^{(1)} = 0$, so there is no correction to $I_{fg}$ at this order. At 2-loop, $Q^{(2)}$ is a diagonal matrix with diagonal elements given by 
\begin{equation}
      Q_{ff}^{(2)} = \sqrt{2}\,(Y_eY_e^{\dagger})_{ff} = \sqrt{2}\,y_f^2 \;,\label{eq:QfgAt2Loop}
\end{equation}
where we use $y_f > 0$. 
Therefore, $\dot I_{fg}$ up to 2-loop using equation  \eqref{eq:FinalResultForDerOfI_inNewNot} is explicitly given by
\begin{equation}
     \frac{\dd I_{fg}}{\dd t} = \frac{2\bigl(y_f^2 - y_g^2\bigr)^2}{(16\piup^2)^2}I_{fg}\;. \label{eq:I_fgUpTo2Loop}
\end{equation}
Specifically,
\begin{subequations}\label{eq:2-loop_RGEs_of_I_fg_2}
    \begin{align}
        \frac{\dd I_{12}}{\dd t} &= \frac{2\bigl(y_e^2 - y_{\mu}^2\bigr)^2}{(16\piup^2)^2}I_{12}\;,\label{eq:I_fgCorrectionTo12} \\[1em]
        \frac{\dd I_{13}}{\dd t} &= \frac{2\bigl(y_e^2 - y_{\tau}^2\bigr)^2}{(16\piup^2)^2}I_{13}\;,\label{eq:I_fgCorrectionTo13} \\[1em]
         \frac{\dd I_{23}}{\dd t} &= \frac{2\bigl(y_{\mu}^2 - y_{\tau}^2\bigr)^2}{(16\piup^2)^2}I_{23} \;.\label{eq:I_fgCorrectionTo23}
    \end{align}
\end{subequations}
Interestingly, these results show that if $I_{fg}$ vanishes at some scale, it will stay zero at all scales. 
Since the coefficients on the right-hand sides of \eqref{eq:2-loop_RGEs_of_I_fg_2} are real, this statement applies separately to the real and imaginary parts of the $I_{fg}$. 
That is, if $\re I_{fg}$ or $\im I_{fg}$ vanishes at some scale, it will remain zero at all scales. 

Given the hierarchy $y_\tau\gg y_\mu\gg y_e$, we see that the corrections of $I_{12}$ are even more suppressed than the \ac{RG} effects on $I_{13}$ and $I_{23}$. 
Even the latter are basically \ac{RG} stable. 
Since $y_\tau\sim10^{-2}$, the coefficient is of the order $10^{-10}$. Multiplying this by $\ln(\Lambda_\mathrm{flavor}/v_{\mathrm{EW}})$ still leads to \ac{RG} effects at most of the order $10^{-8}$.
This means that, for all practical purposes, the $I_{fg}$ are invariant under the renormalization group in the \ac{SM}, and thus not sensitive to the flavor scale $\Lambda_\mathrm{flavor}$. 

Using equations \eqref{eq:I_fgUpTo2Loop}, we can estimate  benchmark values to the quantum corrections, given by
\begin{equation}
    \label{eq:ApproxForDerOfIfg} \Delta I_{fg} \approx \frac{2\bigl(y_f^2 - y_g^2\bigr)^2}{(16\piup^2)^2}I_{fg}\Delta t \; .
\end{equation}
We chose $y_{e} \approx 2 \times 10^{-6}$, $y_{\mu} \approx 5 \times 10^{-4}$, and $y_{\tau}\approx 7 \times 10^{-3}$, and between the energy scale $\mu = 10^3 \ \mathrm{GeV} - 10^6 \ \mathrm{GeV}$ and for $\re I_{fg}$, we obtained $I_{12}\approx -35$, $I_{13}\approx -10$, and $I_{23}\approx 2$ from REAP ~\cite{Antusch:2005gp}. The corrections to $I_{fg}$ using \Cref{eq:ApproxForDerOfIfg} are then given by 
\begin{subequations}
    \begin{align}
        \label{eq:CorrectionToI12}\Delta I_{12} &\approx -1.2\times 10^{-15} \\
        \label{eq:CorrectionToI13}\Delta I_{13} &\approx -1.3\times 10^{-11} \\
        \label{eq:CorrectionToI23}\Delta I_{23} &\approx 2.6\times 10^{-12} 
    \end{align}
\end{subequations}
Clearly, these changes are far smaller than the experimental error bars, including the latest results reported by the JUNO collaboration~\cite{JUNO:2025gmd}. 

We have used a modified version of REAP~\cite{Antusch:2005gp} to verify that, when running the invariants at two loop over a few orders of magnitude, they remain practically unchanged. 
A more detailed numerical study will be presented elsewhere. 

Let us comment on \acp{2HDM}. 
Usually one imposes symmetries to make sure that the charged leptons only couple to one of the Higgs doublets in order to avoid \acp{FCNC}~\cite{Weinberg:1976hu,Glashow:1976nt,Paschos:1976ay}, cf.\ the discussion in \cite{Antusch:2001vn}. 
In these models $y_\tau$ may be of the order unity. 
Even in this case, the corrections \eqref{eq:2-loop_RGEs_of_I_fg_2} remain well below the percent level.

\subsection{Limitations}

In our analysis, we have focused on the case in which the model gives rise to \ac{SM}, \ac{MSSM} or a \ac{2HDM} below its definition scale. 
If there are additional renormalizable couplings that are sensitive to specific lepton flavors, our analysis may no longer apply. 
Studying such scenarios is beyond the scope of this work.

\section{Summary}

Motivated by the analytic properties of certain combinations of the neutrino mass matrix $I_{fg}$ in the context of modular flavor symmetries, we have studied the stability of these expressions under the renormalization group. 
While the $I_{fg}$ receive corrections at the two-loop level, for all practical purposes they remain \ac{RG} invariant in the \ac{SM} and \ac{2HDM}, i.e.\ in the absence of \ac{SUSY}. 
This leads to predictions that are insensitive to the scale $\Lambda_\mathrm{flavor}$ at which the model is defined. 
The conclusions drawn from the analytical properties of the $I_{fg}$ can therefore be confronted to data without the need of a detailed renormalization group analysis. 
In other words, experimental measurements can directly probe high-scale physics. 

\subsection*{Acknowledgments}

We are indebted to Michael Schmidt for help with the REAP package. 
S.P.\ acknowledges support from the APS Bridge program and the GAANN fellowship funded by the U.S. Department of Education. 
\bibliography{RG_invariants}

\providecommand{\href}[2]{#2}\begingroup\raggedright\begin{thebibliography}{10}

\bibitem{Feruglio:2017spp}
F.~Feruglio, {\em {Are neutrino masses modular forms?}},
  \href{http://dx.doi.org/10.1142/9789813238053_0012}{pp.~227--266}.
\newblock 2019.
\newblock \href{http://arxiv.org/abs/1706.08749}{{\ttfamily arXiv:1706.08749
  [hep-ph]}}.

\bibitem{Feruglio:2019ybq}
F.~Feruglio and A.~Romanino, ``{Lepton flavor symmetries},''
  \href{http://dx.doi.org/10.1103/RevModPhys.93.015007}{{\em Rev. Mod. Phys.}
  {\bfseries 93} no.~1, (2021) 015007},
  \href{http://arxiv.org/abs/1912.06028}{{\ttfamily arXiv:1912.06028
  [hep-ph]}}.

\bibitem{Almumin:2022rml}
Y.~Almumin, M.-C. Chen, M.~Cheng, V.~Knapp-Perez, Y.~Li, A.~Mondol,
  S.~Ramos-S{\'a}nchez, M.~Ratz, and S.~Shukla, ``{Neutrino Flavor Model
  Building and the Origins of Flavor and $CP$ Violation: A Snowmass White
  Paper},'' in {\em {Snowmass 2021}}.
\newblock 4, 2022.
\newblock \href{http://arxiv.org/abs/2204.08668}{{\ttfamily arXiv:2204.08668
  [hep-ph]}}.

\bibitem{Kobayashi:2023zzc}
T.~Kobayashi and M.~Tanimoto, ``{Modular flavor symmetric models},''
\newblock 7, 2023.
\newblock \href{http://arxiv.org/abs/2307.03384}{{\ttfamily arXiv:2307.03384
  [hep-ph]}}.

\bibitem{Ding:2023htn}
G.-J. Ding and S.~F. King, ``{Neutrino Mass and Mixing with Modular
  Symmetry},'' \href{http://arxiv.org/abs/2311.09282}{{\ttfamily
  arXiv:2311.09282 [hep-ph]}}.

\bibitem{Ding:2024ozt}
G.-J. Ding and J.~W.~F. Valle, ``{The symmetry approach to quark and lepton
  masses and mixing},''
  \href{http://dx.doi.org/10.1016/j.physrep.2024.12.005}{{\em Phys. Rept.}
  {\bfseries 1109} (2025) 1--105},
  \href{http://arxiv.org/abs/2402.16963}{{\ttfamily arXiv:2402.16963
  [hep-ph]}}.

\bibitem{Feruglio:2025ztj}
F.~Feruglio and S.~Ramos-Sanchez, ``{Quark and lepton masses},''
  \href{http://arxiv.org/abs/2506.20755}{{\ttfamily arXiv:2506.20755
  [hep-ph]}}.

\bibitem{Chen:2024otk}
M.-C. Chen, X.~Li, X.-G. Liu, O.~Medina, and M.~Ratz, ``{Modular invariant
  holomorphic observables},''
  \href{http://dx.doi.org/10.1016/j.physletb.2024.138600}{{\em Phys. Lett. B}
  {\bfseries 852} (2024) 138600},
  \href{http://arxiv.org/abs/2401.04738}{{\ttfamily arXiv:2401.04738
  [hep-ph]}}.

\bibitem{Chang:2002yr}
S.~Chang and T.-K. Kuo, ``{Renormalization invariants of the neutrino mass
  matrix},'' \href{http://dx.doi.org/10.1103/PhysRevD.66.111302}{{\em Phys.
  Rev. D} {\bfseries 66} (2002) 111302},
  \href{http://arxiv.org/abs/hep-ph/0205147}{{\ttfamily arXiv:hep-ph/0205147}}.

\bibitem{Qu:2024rns}
B.-Y. Qu and G.-J. Ding, ``{Non-holomorphic modular flavor symmetry},''
  \href{http://dx.doi.org/10.1007/JHEP08(2024)136}{{\em JHEP} {\bfseries 08}
  (2024) 136}, \href{http://arxiv.org/abs/2406.02527}{{\ttfamily
  arXiv:2406.02527 [hep-ph]}}.

\bibitem{Ding:2024inn}
G.-J. Ding, J.-N. Lu, S.~T. Petcov, and B.-Y. Qu, ``{Non-holomorphic modular
  S$_{4}$ lepton flavour models},''
  \href{http://dx.doi.org/10.1007/JHEP01(2025)191}{{\em JHEP} {\bfseries 01}
  (2025) 191}, \href{http://arxiv.org/abs/2408.15988}{{\ttfamily
  arXiv:2408.15988 [hep-ph]}}.

\bibitem{Qu:2025ddz}
B.-Y. Qu, J.-N. Lu, and G.-J. Ding, ``{Non-holomorphic modular flavor symmetry
  and odd weight polyharmonic Maa{\ss} form},''
  \href{http://arxiv.org/abs/2506.19822}{{\ttfamily arXiv:2506.19822
  [hep-ph]}}.

\bibitem{Esteban:2024eli}
I.~Esteban, M.~C. Gonzalez-Garcia, M.~Maltoni, I.~Martinez-Soler, J.~P.
  Pinheiro, and T.~Schwetz, ``{NuFit-6.0: updated global analysis of
  three-flavor neutrino oscillations},''
  \href{http://dx.doi.org/10.1007/JHEP12(2024)216}{{\em JHEP} {\bfseries 12}
  (2024) 216}, \href{http://arxiv.org/abs/2410.05380}{{\ttfamily
  arXiv:2410.05380 [hep-ph]}}.

\bibitem{JUNO:2025gmd}
{\bfseries JUNO} Collaboration, A.~Abusleme {\em et al.}, ``{First measurement
  of reactor neutrino oscillations at JUNO},''
  \href{http://arxiv.org/abs/2511.14593}{{\ttfamily arXiv:2511.14593
  [hep-ex]}}.

\bibitem{Capozzi:2025ovi}
F.~Capozzi, E.~Lisi, F.~Marcone, A.~Marrone, and A.~Palazzo, ``{Updated bounds
  on the (1,2) neutrino oscillation parameters after first JUNO results},''
  \href{http://arxiv.org/abs/2511.21650}{{\ttfamily arXiv:2511.21650
  [hep-ph]}}.

\bibitem{Haba:1999ca}
N.~Haba, Y.~Matsui, N.~Okamura, and M.~Sugiura, ``{Energy scale dependence of
  the lepton flavor mixing matrix},''
  \href{http://dx.doi.org/10.1007/s100520050605}{{\em Eur. Phys. J. C}
  {\bfseries 10} (1999) 677--680},
  \href{http://arxiv.org/abs/hep-ph/9904292}{{\ttfamily arXiv:hep-ph/9904292}}.

\bibitem{Antusch:2001ck}
S.~Antusch, M.~Drees, J.~Kersten, M.~Lindner, and M.~Ratz, ``{Neutrino mass
  operator renormalization revisited},''
  \href{http://dx.doi.org/10.1016/S0370-2693(01)01127-3}{{\em Phys. Lett. B}
  {\bfseries 519} (2001) 238--242},
  \href{http://arxiv.org/abs/hep-ph/0108005}{{\ttfamily arXiv:hep-ph/0108005}}.

\bibitem{Antusch:2001vn}
S.~Antusch, M.~Drees, J.~Kersten, M.~Lindner, and M.~Ratz, ``{Neutrino mass
  operator renormalization in two Higgs doublet models and the MSSM},''
  \href{http://dx.doi.org/10.1016/S0370-2693(01)01414-9}{{\em Phys. Lett. B}
  {\bfseries 525} (2002) 130--134},
  \href{http://arxiv.org/abs/hep-ph/0110366}{{\ttfamily arXiv:hep-ph/0110366}}.

\bibitem{Chankowski:1993tx}
P.~H. Chankowski and Z.~Pluciennik, ``{Renormalization group equations for
  seesaw neutrino masses},''
  \href{http://dx.doi.org/10.1016/0370-2693(93)90330-K}{{\em Phys. Lett. B}
  {\bfseries 316} (1993) 312--317},
  \href{http://arxiv.org/abs/hep-ph/9306333}{{\ttfamily arXiv:hep-ph/9306333}}.

\bibitem{Babu:1993qv}
K.~S. Babu, C.~N. Leung, and J.~T. Pantaleone, ``{Renormalization of the
  neutrino mass operator},''
  \href{http://dx.doi.org/10.1016/0370-2693(93)90801-N}{{\em Phys. Lett. B}
  {\bfseries 319} (1993) 191--198},
  \href{http://arxiv.org/abs/hep-ph/9309223}{{\ttfamily arXiv:hep-ph/9309223}}.

\bibitem{Xing:2007fb}
Z.-z. Xing, H.~Zhang, and S.~Zhou, ``{Updated Values of Running Quark and
  Lepton Masses},'' \href{http://dx.doi.org/10.1103/PhysRevD.77.113016}{{\em
  Phys. Rev. D} {\bfseries 77} (2008) 113016},
  \href{http://arxiv.org/abs/0712.1419}{{\ttfamily arXiv:0712.1419 [hep-ph]}}.

\bibitem{Xing:2020ijf}
Z.-z. Xing, ``{Flavor structures of charged fermions and massive neutrinos},''
  \href{http://dx.doi.org/10.1016/j.physrep.2020.02.001}{{\em Phys. Rept.}
  {\bfseries 854} (2020) 1--147},
  \href{http://arxiv.org/abs/1909.09610}{{\ttfamily arXiv:1909.09610
  [hep-ph]}}.

\bibitem{Machacek:1984zw}
M.~E. Machacek and M.~T. Vaughn, ``{Two Loop Renormalization Group Equations in
  a General Quantum Field Theory. 3. Scalar Quartic Couplings},''
  \href{http://dx.doi.org/10.1016/0550-3213(85)90040-9}{{\em Nucl. Phys. B}
  {\bfseries 249} (1985) 70--92}.

\bibitem{Machacek:1983fi}
M.~E. Machacek and M.~T. Vaughn, ``{Two Loop Renormalization Group Equations in
  a General Quantum Field Theory. 2. Yukawa Couplings},''
  \href{http://dx.doi.org/10.1016/0550-3213(84)90533-9}{{\em Nucl. Phys. B}
  {\bfseries 236} (1984) 221--232}.

\bibitem{Luo:2002ti}
M.-x. Luo, H.-w. Wang, and Y.~Xiao, ``{Two loop renormalization group equations
  in general gauge field theories},''
  \href{http://dx.doi.org/10.1103/PhysRevD.67.065019}{{\em Phys. Rev. D}
  {\bfseries 67} (2003) 065019},
  \href{http://arxiv.org/abs/hep-ph/0211440}{{\ttfamily arXiv:hep-ph/0211440}}.

\bibitem{Schienbein:2018fsw}
I.~Schienbein, F.~Staub, T.~Steudtner, and K.~Svirina, ``{Revisiting RGEs for
  general gauge theories},''
  \href{http://dx.doi.org/10.1016/j.nuclphysb.2018.12.001}{{\em Nucl. Phys. B}
  {\bfseries 939} (2019) 1--48},
  \href{http://arxiv.org/abs/1809.06797}{{\ttfamily arXiv:1809.06797
  [hep-ph]}}. [Erratum: Nucl.Phys.B 966, 115339 (2021)].

\bibitem{Ibarra:2024tpt}
A.~Ibarra, N.~Leister, and D.~Zhang, ``{Complete two-loop renormalization group
  equation of the Weinberg operator},''
  \href{http://dx.doi.org/10.1007/JHEP03(2025)214}{{\em JHEP} {\bfseries 03}
  (2025) 214}, \href{http://arxiv.org/abs/2411.08011}{{\ttfamily
  arXiv:2411.08011 [hep-ph]}}.

\bibitem{Antusch:2005gp}
S.~Antusch, J.~Kersten, M.~Lindner, M.~Ratz, and M.~A. Schmidt, ``{Running
  neutrino mass parameters in see-saw scenarios},''
  \href{http://dx.doi.org/10.1088/1126-6708/2005/03/024}{{\em JHEP} {\bfseries
  03} (2005) 024}, \href{http://arxiv.org/abs/hep-ph/0501272}{{\ttfamily
  arXiv:hep-ph/0501272}}.

\bibitem{Weinberg:1976hu}
S.~Weinberg, ``{Gauge Theory of CP Violation},''
  \href{http://dx.doi.org/10.1103/PhysRevLett.37.657}{{\em Phys. Rev. Lett.}
  {\bfseries 37} (1976) 657}.

\bibitem{Glashow:1976nt}
S.~L. Glashow and S.~Weinberg, ``{Natural Conservation Laws for Neutral
  Currents},'' \href{http://dx.doi.org/10.1103/PhysRevD.15.1958}{{\em Phys.
  Rev. D} {\bfseries 15} (1977) 1958}.

\bibitem{Paschos:1976ay}
E.~A. Paschos, ``{Diagonal Neutral Currents},''
  \href{http://dx.doi.org/10.1103/PhysRevD.15.1966}{{\em Phys. Rev. D}
  {\bfseries 15} (1977) 1966}.

\end{thebibliography}\endgroup
\bibliographystyle{utphys}

\end{document}